\documentclass[twocolumn]{aa}  
\usepackage{longtablefn}
\usepackage{graphicx}
\usepackage{lscape}
\usepackage{txfonts}
\usepackage{natbib}
\bibpunct{(}{)}{;}{a}{}{,}

\begin{document}
   \title{Deep near-IR variability survey of pre-main-sequence stars in $\rho$~Ophiuchi}
   \author{C. Alves de Oliveira\inst{1} \and M. Casali\inst{1}     }
   \offprints{C. Alves de Oliveira}
   \institute{European Southern Observatory (ESO), Karl-Schwarzschild-Strasse 2
D-85748 Garching bei Muenchen,  Germany \\
              \email{coliveir@eso.org}}
   \date{Accepted for publication in A$\&$A}

\abstract
{Variability is a common characteristic of pre-main-sequence stars (PMS). Near-IR variability surveys of young stellar objects (YSOs) can probe stellar and circumstellar environments and provide information about the dynamics of the on going magnetic and accretion processes. Furthermore, variability can be used as a tool to uncover new cluster members in star formation regions.}
{We hope to achieve the deepest near-IR variability study of YSOs targeting the $\rho$~Ophiuchi cluster.}
{Fourteen epochs of observations were obtained with the Wide Field Camera (WFCAM) at the UKIRT telescope scheduled in a manner that allowed the study of variability on timescales of days, months, and years. Statistical tools, such as the multi-band cross correlation index and the reduced chi-square, were used to disentangle signals of variability from noise. Variability characteristics are compared to existing models of YSOs in order to relate them to physical processes, and then used to select new candidate members of this star-forming region. }
{Variability in the near-IR is found to be present in 41\% of the known population of $\rho$~Ophiuchi recovered in our sample. The behaviours shown are several and can be associated with the existence of spots on the stellar surface, variations in circumstellar extinction, or changes in the geometry of an accretion disc. Using variability, a new population of objects has been uncovered that is believed to be part of the $\rho$~Ophiuchi cluster.}
    {}
\keywords{Stars: pre-main-sequence -- 
	 stars: low-mass, brown dwarfs --
         stars: activity --
         stars: variables: general}

\titlerunning{Near-IR variability survey of PMS stars in $\rho$~Ophiuchi} 
\authorrunning{C. Alves de Oliveira \& M. Casali} 
\maketitle
%________________________________________________________________

\section{Introduction}
Young stars have been known to be variable since \citet{Joy1945} described the irregular behaviour of T Tauri stars. Photometric variability is thought to originate from several mechanisms related to magnetic fields, accretion discs, and circumstellar extinction \citep{Herbst1994}. Optical variability has been used as an excellent tool for characterising stellar and circumstellar environments of pre-main-sequence stars \citep[e.g.,][]{Grankin2007,Grankin2008} and brown dwarfs \citep[e.g.,][]{Caballero2004,Scholz2005}. The mechanisms causing optical variability in YSOs are also thought to be responsible for the occurrence of photometric variations in the near-IR, especially suitable for probing phenomena which take place in the circumstellar temperature regime \citep[see][and references therein]{Eiroa2002}. Furthermore, \citet{Kaas1999} has shown that IR variability is a very useful tool for sorting the cluster population from the background field. For instance, it has the advantage of identifying young sources that do not show IR-excess or have small H$\alpha$ emission, as is the case for weak-line T~Tauri stars (WTTS), but are magnetically active \citep{Grankin2008}. The use of this relatively new technique should contribute significantly to obtain a full census of the young objects in star formation regions, one of the most important goals of galactic star formation studies.

At a distance of 119$\pm$6~pc \citep{Lombardi2008}, the $\rho$~Ophiuchi cloud core is one of the nearest star forming regions, and therefore widely studied. Despite the proximity, few stars are optically observable due to the high visual extinction in the core, estimated to be 50-100 magnitudes \citep{Wilking1983}. That is also the case for most Myr old clusters, where much of the interesting population remains visibly obscured, and must be studied with IR observations. However, with many nearby star forming regions extending over degree scales and IR cameras usually having fields of only arc minutes, obtaining a single image of a region such as $\rho$~Ophiuchi has been a major project until now, let alone attempting multi-epoch studies. This has changed with 2MASS which allows large area, though rather shallow, observations to be made, and near-IR variability studies to become possible \citep{Carpenter2001,Carpenter2002}. In a 2MASS \emph{JHK} study centred near the Trapezium region of the Orion Nebula Cluster, where most star formation is thought to have occurred between 0.3 and 2 Myr ago \citep{Ali1995,Hillenbrand1997}, \citet{Carpenter2001} found that approximately 45\% of young stars were variable.

More recently, the WFCAM IR imager \citep{Casali2007} has become operational on the UK Infrared Telescope (UKIRT). The large field of view ($\sim$0.8 deg$^{2}$ in four exposures) of this instrument has made deep variability studies possible for the first time in the IR. This paper presents the results of such a study, with multi-epoch \emph{H} and \emph{K} observations of the $\rho$~Ophiuchi cluster used to search for variability on timescales of days, months and years. The study is enhanced by including recently released IRAC/Spitzer data on $\rho$~Ophiuchi \citep{Evans2005}. In Sect.~2, the observations and reductions for the large data sample are described. Sect.~3 explains the methods used in the search for variability together with the results of the analysis. These are discussed and explained in a physical context in Sect.~4, in an attempt to relate the variability behaviours observed to the possible physical causes. Conclusions and future prospects are given in Sect.~5. \\

%__________________________________________________________________
\section{Observations}

The WFCAM on the UKIRT is a survey instrument developed at the UK Astronomical Technology Centre (UK ATC), primarily for the UKIRT Infrared Deep Sky Survey (UKIDSS) \citep{Lawrence2007}. WFCAM is a wide field imaging camera operating in the near infrared from 0.83~$\mu$m to 2.37~$\mu$m in up to eight filters, including \emph{ZY~JHK} \citep{Hewett2006}. The camera uses four Rockwell Hawaii-II 2048~x~2048 18~$\mu$m-pixel array detectors with a pixel scale of $0.4\arcsec$. The four detectors are arranged in a 2~x~2 pattern and are separated by $94$\% of a detector width, for which four exposures are needed to survey a contiguous area of $\sim$0.8 deg$^{2}$ \citep{Casali2007}.

The data presented amount to a total of 14 epochs in the $\rho$ Ophiuchi cluster. The observed region consists of four exposures which produce a final tile of $\sim$0.8 deg$^{2}$ centred on the cloud's core. Each observing block was done in \emph{H} and \emph{K} bands where completeness limits of magnitudes 19.0 and 18.0, respectively, are achieved. Table~\ref{table:1} shows the central position (right ascension and declination) for each of the four exposures. The same region in the sky was observed randomly over scales of days, weeks, months and year sampling periods.  Table~\ref{table:2} provides a list of the nights in which the observations occurred. 

%__________________________________________________ Table of coordinates Obs.
\begin{table}
\caption{Coordinates of the observations.}           % title of Table
\label{table:1}      % is used to refer this table in the text
\centering                          % used for centering table
\begin{tabular}{c c c}        % centered columns (4 columns)
\hline            % inserts double horizontal lines
\hline
Pointing & RA & Dec. \\
\hline                        % inserts single horizontal line
1 & 16 27 20.0 & $-24$ 19 20     \\ 
2 & 16 27 20.0 & $-24$ 32 34     \\
3 & 16 28 18.2 & $-24$ 29 00     \\
4 & 16 28 18.2 & $-24$ 15 48     \\
 \hline                                   %inserts single line
\end{tabular}
\end{table}
%______________________________________________   

%______________________________________________   Table of dates of Obs. 
\begin{table}
\begin{minipage}[t]{\columnwidth}
\caption{Log of the observations.}           % title of Table
\label{table:2}      % is used to refer this table in the text
\centering                          % used for centering table
\renewcommand{\footnoterule}{}
\begin{tabular}{c c c c}
\hline
\hline
UT date & May & June  & July \\
\hline
 2005   & 11, 18, 21, 29 & 04, 07, 14 & ... \\
\hline
 2006 	& 24, 28\footnote{Observations taken only for $\emph{H}$ band.} & 03, 14, 23 & 04, 10   \\
\hline
\end{tabular}
\end{minipage}
\end{table}
%______________________________________________   

To complement the study, Spitzer data from the c2d legacy project \citep{Evans2003} have been included. The Ophiuchus molecular cloud has been mapped with IRAC \citep{Fazio2004} in the 3.6, 4.5, 5.8 and 8.0~$\mu$m bands, over a region of 8.0~deg$^{2}$, which encompasses the WFCAM field. The data have been retrieved from the preliminary c2d IRAC point-source catalogues of the third data delivery \citep{Evans2005}.

\subsection{Data processing}

WFCAM data are processed and archived by the VISTA Data Flow System Project, a collaboration between Queen Mary University of London, the Institute of Astronomy of the University of Cambridge and the Institute for Astronomy of the University of Edinburgh (Irwin et al., \emph{in preparation}). The pipeline flat-fields the data, subtracts the counts from the background sky, detects and parameterises objects and performs the photometric and astrometric calibrations. The products from the pipeline are a set of reduced uncalibrated individual exposures, and photometrically and astrometrically calibrated stacked frames, as well as catalogues of sources detected in the frames and ingested into the WFCAM Science Archive, described by \citet{Hambly2008}. A summary description of the WFCAM pipeline and science archive is given by \citet{Lawrence2007}.

Given the expected very low amplitudes of variation in young stars, a few tenths or less of a magnitude in the near-IR \citep{Carpenter2001}, it is important to be aware of common problems intrinsic to the instrument and type of observations. To a large extent, many of these problems are eliminated by the pipeline during reduction and calibration. However, others do persist giving rise to false detections which show up in the catalogues. Although they represent a small fraction of the dataset, they need to be filtered out. A routine was written in IDL to implement an algorithm which finds and removes these spurious sources and implements the necessary constraints to ensure a reliable final source list. These are briefly explained below.
\begin{itemize}
	\item Overlapping regions: the algorithm accounted for sources in the overlapping regions between detectors which have multiple entries in the catalogue, keeping the source with the largest distance to the detector edge.
	\item Magnitude limits: sets a cut-off limit at the brighter end (magnitudes 11.0 and 10.0 for \emph{H} and \emph{K}, respectively) due to saturation. At the fainter end, magnitudes 19.0 and 18.0 for \emph{H} and \emph{K}, respectively, mark the completeness limits which are above the 10$\sigma$ level (see Table~\ref{table:3} for the average errors).
	
%__________________________________________________ Table of accuracy

\begin{table}
\caption{Photometric accuracy.}           % title of Table
\label{table:3}      % is used to refer this table in the text
\centering                          % used for centering table
\begin{tabular}{c c c}        % centered columns (4 columns)
\hline            % inserts double horizontal lines
\hline
Magnitude & \multicolumn{2}{c}{Photometric Accuracy}  \\
~ & \emph{H}  & \emph{K} \\
\hline                        % inserts single horizontal line
10-11 &  ...  & 0.01 \\ 
11-12 &  0.01 & 0.01 \\
12-13 &  0.01 & 0.01 \\
13-14 &  0.01 & 0.01 \\
14-15 &  0.01 & 0.01 \\
15-16 &  0.01 & 0.02 \\
16-17 &  0.02 & 0.05 \\ 
17-18 &  0.04 & 0.07 \\
18-19 &  0.05 & ...  \\
\hline                                   %inserts single line
\end{tabular}
\end{table}
%______________________________________________   

	\item Object classification: objects flagged by the pipeline as stars were kept as well as objects which were classified as galaxies but showed no difference in magnitude between different apertures. The motivation was to keep some nebulous YSOs while setting the criteria strictly enough to avoid galaxy contamination.
	\item Cross-talk: electronic cross-talk occurs between the detector channels within a quadrant. It produces a sequence of spurious images in some or all the other $7$ channels \citep{Dye2006}. The algorithm removed sources located at up to seven multiples of 128~pix from bright stars.
	\item Diffraction patterns: sources in the surroundings of saturated stars were removed since their photometry is not reliable. The radius of rejection was adjusted according to the brightness of the star. 
	\item Field edges: sources located on the outer field edge were removed since they were found to often have unreliable photometry.	
\end{itemize}

Furthermore, the \emph{K} band images from the observations made on 28 May 2006 were discarded since they were of lower quality due to technical problems.

\begin{figure}
   \centering
  \resizebox{\hsize}{!}{\includegraphics{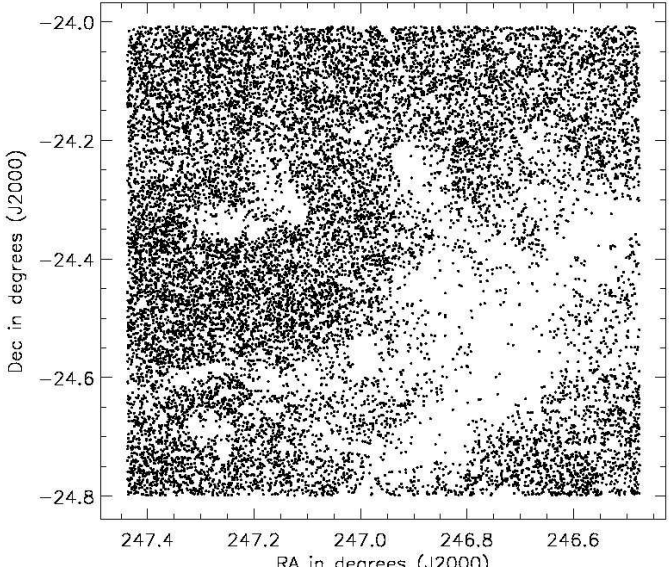}}
      \caption{Tile coverage in the $\rho$ Ophiuchi cluster. The total area covered is 0.8 deg$^{2}$. The spatial distribution of the detected sources reflects the cloud extinction. The visual extinction in the core is estimated to be 50-100 magnitudes \citep{Wilking1983}.}
         \label{figure:1}
   \end{figure}

The catalogues for each night were merged for the two years separately. Sources were matched by requiring their position to be within an angular separation of $1.0\arcsec$, since the position offsets found range from $0.1\arcsec$ to $~0.8\arcsec$. It was further required that each source had at least four epochs with detections in each filter. This criterion was implemented in order to reduce the probability of false transient detections and to reduce the effects of close double stars which are not resolved in all the nights, depending on seeing conditions. In total, 15\,191 sources were detected across the two years of observations (Fig.~\ref{figure:1}).

The internal accuracy of the photometry between nights was further improved by measuring and removing offsets,  which can arise for several reasons. The photometric calibration of WFCAM data is done with 2MASS stars in the observed frames \citep{Dye2006,Warren2007}. However, some of these stars could be variable, which would have an impact on the calibration. Furthermore, for regions in the sky with high extinction, as is the case for the central core of $\rho$ Ophiuchi, the number of available 2MASS stars is reduced, which limits the photometric accuracy. Lastly, photometric offsets can also arise from small problems in flatfielding and sky subtraction from night to night. The removal of these offsets was done separately per detector and for each of the four sky pointings. Stars with magnitudes between 12.0 and 16.0 for \emph{K} band and 13.0 to 17.0 in \emph{H} band were used for the calculations, assuming that their magnitudes, in the median, do not change in time. The final offset for one detector at a given sky pointing was taken to be the average value of the differences in magnitude between the several nights and a reference night. This offset value, up to 0.03 magnitudes, was then removed from all the sources detected in that particular detector and sky pointing.

Spitzer counterparts were obtained from the c2d IRAC point-source catalogues of the third data delivery, applying the recommended criteria to select highly reliable samples, i.e. only considering sources which are not extended, as estimated by the source extraction, and with a quality flag of A or B in the four IRAC bands, which corresponds to a signal$-$to$-$noise ratio $\ga$10 and $\ga$7, respectively \citep[see][for a detailed explanation]{Evans2005}. Spitzer fluxes were converted to magnitudes using the IRAC zero magnitude flux densities defined by \citet{Reach2005}.

%__________________________________________________________________

\section{Results}
     
\subsection{List of near-IR variable stars}     
     
A variable star is one which displays photometric characteristics incompatible with photometric errors. The large sample being analysed required the use of statistical tools which can quantitatively estimate the probability that the detected photometric variations are intrinsic to the star or its circumstellar material.

The first tool used to detect variability was the reduced chi-square $\chi^{2}(\chi^{2}_{\nu})$ of the magnitudes, computed using:

\begin{equation}
\chi^{2}_{\nu} = \frac{1}{\nu} \sum^{N}_{i=1} \frac{(mag_{i} - \bar{mag})^2}{\sigma^{2}_{i}} ,
\end{equation}

where \emph{$\nu$} is the number of degrees of freedom, \emph{N} is the number of measurements and \emph{$\sigma_{i}$} is the photometric uncertainty. In this case, the statistic represents the probability that the variations result from Gaussian noise. Therefore, the reduced chi-square index entirely relies on the assumption that the noise is Gaussian, which is only approximately the case in any real set of observations in which there is variable seeing, bad pixels, detector imperfections, etc. It is also very susceptible to outlier points and does not make use of correlated changes in multiband photometry, further limiting its use. 

The second tool used takes advantage of the temporal coherence of the star$'$s light curve and also the correlation across  \emph{H} and \emph{K} bands, allowing for true signals to be extracted from within the noise. The cross-correlation index (CCI) relies on simple correlation coefficients statistics and was first proposed for use in the search of variable stars by \citet{Welch1993}. This method, which assumes that photometric errors for two different photometry lists should be uncorrelated, is effective as long as the difference in time between observations in each frame pair (at a given epoch) is small compared with the variation period. The index is defined as:

\begin{equation}
CCI = \sqrt{\frac{1}{N(N-1)}} \sum^{N}_{i=1} \left(\frac{mag^{H}_{i} - \bar{mag}}{\sigma_{H,i}}\right) \left(\frac{mag^{K}_{i} - \bar{mag}}{\sigma_{K,i}}\right) ,
\end{equation}

where \emph{N} is the number of measurements and \emph{$\sigma_{H,i}$}, \emph{$\sigma_{K,i}$} are the photometric uncertainties.

A third index (CI$_H$, CI$_K$) was computed for each band separately which takes advantage of the temporal coherence of the light curve and can be used to detect long term variations in a single band. It is defined as

\begin{equation}
CI_{K} = \sqrt{\frac{1}{N(N-1)}} \sum^{N}_{i=1} \left(\frac{mag^{K}_{i} - \bar{mag}}{\sigma_{K,i}}\right) \left(\frac{mag^{K}_{i+1} - \bar{mag}}{\sigma_{K,i+1}}\right) ,
\end{equation} 

for \emph{K} band, and similarly for the \emph{H} band.

\begin{figure}
   \centering
   \resizebox{\hsize}{!}{\includegraphics{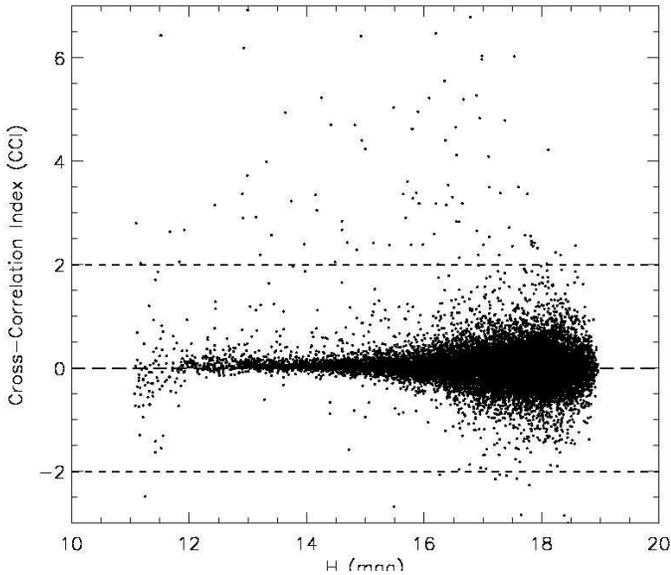}}
      \caption{Cross correlation index (CCI). Variable stars show positive values of the CCI greater than 2.}
         \label{figure:2}
   \end{figure}

For each star, the $\chi^{2}_{\nu}$, CCI, CI$_H$ and CI$_K$, were computed for each year separately. A star was classified as variable when it met one of the following criteria: 
\begin{itemize}
	\item $\chi^{2}_{\nu}$~$>$~17.3 in one or two bands, meaning only 1 to 2 (to be conservative) false detections in the total sample are expected

	\item  CCI, CI$_H$ or CI$_K$~$>$~2.0. Numerical simulations have been used to study the behaviour of the correlation indices. For ~16000 stars (approximately the size of the dataset), the distribution of values for the CCI due to random chance over 7 epochs is symmetric about zero and has a maximum value of 1.8. The simulations for the individual band correlation indices showed the same behaviour, with a maximum value of the indices due to random chance of 1.9. Figure~\ref{figure:2} shows the CCI as a function of \emph{H} magnitude. Non-variable stars produce a probability distribution centred about zero due to noise while true variables show positive values of the CCI greater than 2 (to be conservative). All the correlation indices showed a similar behaviour. The width of the distribution of non-variable stars depends on the total number of epochs used in the calculation of the correlation indices, and so the threshold was scaled accordingly (CCI, CI$_H$ or CI$_K$~$>$~2, 2.16, 2.37, 2.65 for a star with detections in 7, 6, 5, or 4 epochs, respectively).
	
\end{itemize}

\addtocounter{table}{1}

The application of these criteria led to a raw list of 235 candidates for variability. These were visually examined and 59 objects were rejected, primarily because they turned out to be close double stars which fell inside the aperture, showing fluctuations in magnitude which originated from differences in seeing across the different epochs. Furthermore, faint objects (\emph{K}$>$15.5) which showed variability characteristics consistent with the properties of active field M dwarfs were also removed, since they are more likely to be affected by field contamination. Active low mass M-dwarf (dMe) stars have strong surface magnetic fields \citep{Johns1996} and show significant levels of coronal activity with maximum amplitude variations in the optical of $<$0.5 magnitudes \citep{Bondar2002}. Assuming the same value of amplitude variation for IR variability (to be conservative), a simple starspot model \citep{Vrba1986} predicts a maximum colour variation \emph{H}$-$\emph{K} of 0.015 magnitudes. All variables fainter than 15.5 in \emph{K}--band, magnitude amplitude variations less than 0.5 and colour variations less than 0.015, with a 3$\sigma$ confidence, were therefore not considered in the final list of candidate members of $\rho$~Ophiuchi since they were consistent with being active M~dwarfs. This sample could, however, contain genuine cluster members, and so a table with the main properties has been included in an appendix.\\ 

The final list contains 137 variable stars. Table~\ref{stars} shows the photometric properties for each variable star, such as the coordinates, the average \emph{H} and \emph{K} magnitudes, the peak-to-peak amplitude for \emph{H}, \emph{K}, and \emph{H}$-$\emph{K}, the slope in the variability colour-magnitude diagram (Sect.~3.4), the SED class as determined from IRAC/Spitzer colour-colour diagrams (Sect.~3.5), several variability flags (Sect.~3.4), and references for the known members of $\rho$~Ophiuchi (Sect.~3.2). 

\subsection{Known Population of the Ophiuchus Molecular Cloud}

Many imaging and spectroscopic studies have been done on the Ophiuchus molecular cloud across the spectrum, mainly at near- and mid-infrared wavelengths, \citep[e.g.,][]{Vrba1975,Elias1978,Wilking1989,Comeron1993,Luhman1999,Bontemps2001,Allen2002,Natta2006}, but also in X-rays \citep{Gagne2004,Ozawa2005} and in the optical \citep[e.g.,][]{Martin1998,Wilking2005}. The WFCAM catalogue was merged with several catalogues of $\rho$~Ophiuchi members \citep[namely from the following studies,][]{Comeron1993,Bontemps2001,Gagne2004,Ozawa2005,Wilking2005}, which were chosen based on their accuracy to assess membership, and also to better match the magnitude limits and spatial distribution of the WFCAM observations. WFCAM counter-parts are found for 128 previously known members of $\rho$~Ophiuchi. Of the sources not merged with these catalogues, many are present in source tables of other surveys in the literature, but they did not posses characteristics which enable them to be classified as young cloud members rather than background stars or galaxies. A significant fraction of the $\rho$~Ophiuchi population is not recovered simply because the objects are too bright, and therefore heavily saturated in the WFCAM images. Also the very extended objects are not present in the final catalogues since, to avoid galaxy contamination, this type of object was rejected. From the 128 known members with WFCAM counter-parts, 41\% are variable stars, according to the previous definitions. In the sections that follow, variable stars which have been confirmed as members of the cloud by these studies are referred to as \emph{members}. Variables which have not been confirmed as members are referred to as \emph{candidate members}. 

\subsection{Magnitudes and colours of the variable stars}  

\begin{figure}
   \centering
   \resizebox{\hsize}{!}{\includegraphics{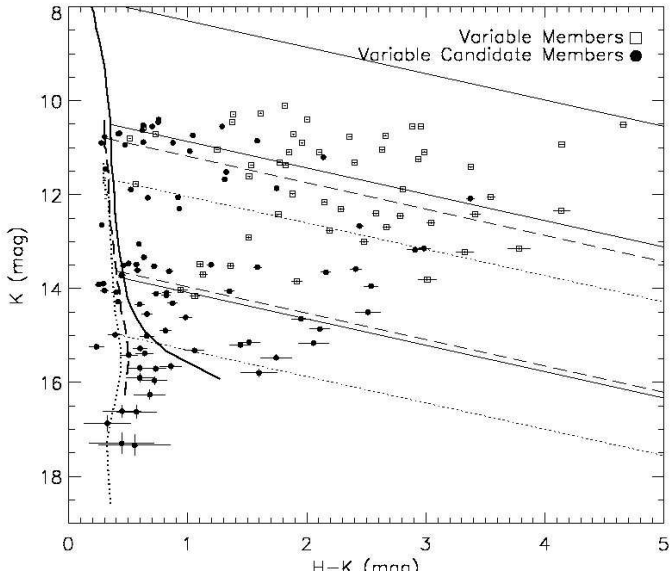}}
      \caption{Colour-magnitude diagram for variable members (open squares) and variable candidate members (filled circles) of $\rho$~Ophiuchi. Solid, dashed and dotted lines show the isochrones at 119 pc of the Lyon group (see text for detailed explanation). Lines parallel to the redening vector \citep{Rieke1985} mark the separation between stars, brown dwarfs and planetary mass candidates for the different isochrones.}
         \label{figure:3}
   \end{figure}

The colours and magnitudes of the variable stars are an important clue in investigating the youth and masses of these objects. The colour-magnitude diagram, \emph{K} versus \emph{H}-\emph{K}, shown in Fig.~\ref{figure:3}, displays the average magnitude and colour of the variable members of $\rho$~Ophiuchi (open squares) and the candidate members (filled circles). The age of objects in the core of $\rho$~Ophiuchi has been found to be 0.3Myr \citep{Greene1995,Luhman1999} and an extensive H$\alpha$ survey by \citep{Wilking2005} derived a median age for of 2.1 Myr, with some members up to 3~Myr, for more widely distributed members. The curves in the diagram  show the model evolutionary tracks of the Lyon group \citep{Baraffe1998,Chabrier2000,Baraffe2003} for a cluster at a distance of ~119 pc \citep{Lombardi2008}. The solid curve shows the theoretical 1~Myr isochrone, for low mass stars ranging from 0.5~M$_{\sun}$ down to 0.003~M$_{\sun}$ (3M$_{\emph{Jup}}$). The isochrone is a combination of the NextGen and the DUSTY isochrones \citep{Baraffe1998, Chabrier2000}. The dashed and dotted curves show the COND isochrones \citep{Baraffe2003} for 1 and 5~Myr, respectively, down to  3M$_{\emph{Jup}}$. The lines parallel to the redening vector \citep{Rieke1985} mark the separation between stars, brown dwarfs (M$<$0.075M$_{\sun}$) and planetary mass candidates (M$<$0.012M$_{\sun}$) for the different isochrones. The models indicate that the brighter candidates are above the hydrogen burning limit, but the faint sources would extend to very low masses. If an age of ~1~Myr is considered, the faintest variable objects would have just a few Jupiter masses. However, when an older age is considered (dotted line), the objects would become more massive. \citet{Wilking2005} found a distributed population in Ophiuchus which is significantly older than that in the more highly extincted cloud core, with ages up to 3~Myr, so it is plausible that part of the variable objects are older than the core population. Although the CMD gives a good indication of the approximate masses of the variable objects, a spectroscopic follow-up is needed to investigate their nature.

\subsection{Characteristics of the variability}

\begin{figure}
   \centering
   \resizebox{\hsize}{!}{\includegraphics{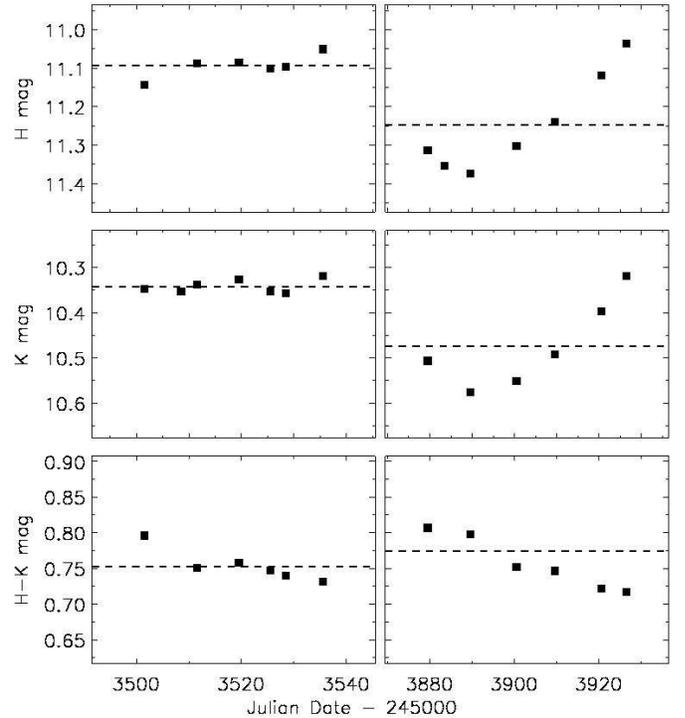}}
   \caption{AOC~J162814.77-242322.5; variable candidate member star with no significant variations in the first year and large variations in the second year. The panels show the \emph{H}, \emph{K}, \emph{H}$-$\emph{K} light curves. The left panels show the data points from 2005 May-June, and the right panels the photometry from 2006 May-July. The dashed lines show the mean values.}
         \label{figure:4}
   \end{figure}

The timescales of variability which can be studied with this dataset can be divided into two groups: intermediate and long time scales. The first applies to the stars which only show variability in one of the two years of observations, and are likely to represent physical phenomena which have a maximum duration of a few months. The long term variables are stars that show variability behaviour across both years, which could be caused either by a single mechanism or be the result of repeated shorter term variability phenomena. In the list of member variables, 77\% of the stars show variability over the two years, while 23\% appear to be variable only in the first or second year. The values for the candidate members are slightly different, where 52\% of the objects are variable in both years and the remaining 48\% show variability in only one of the years of observations. Figure~\ref{figure:4} is an example of a candidate member of $\rho$~Ophiuchi, which shows no significant variations in the first year but large variations in the second year. Given the sparse cadence of the observations, a search for periods has not been attempted. \\

\begin{figure*}
   \centering
   \resizebox{\hsize}{!}{\includegraphics{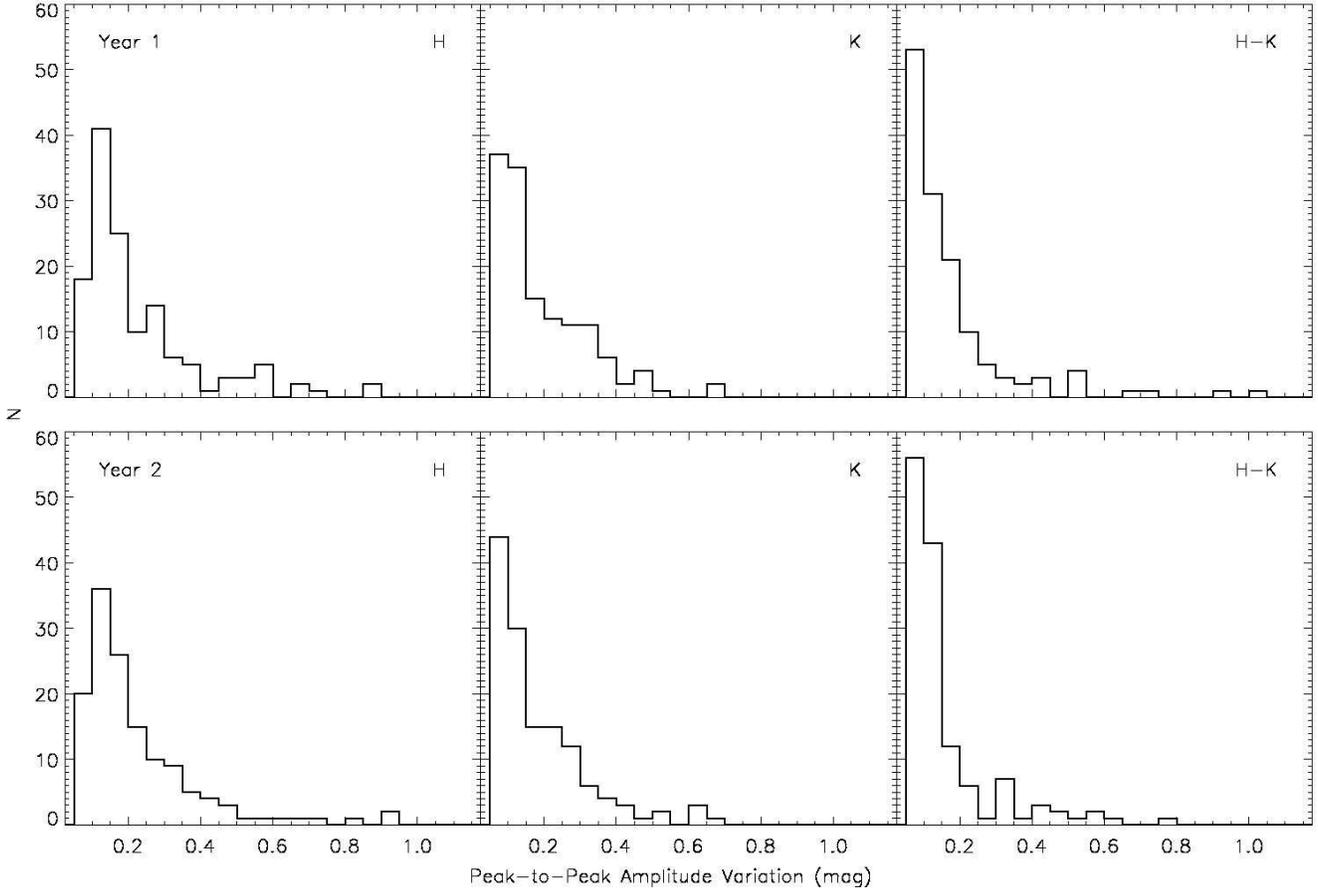}}
      \caption{Histograms of the peak-to-peak magnitude amplitude of the variations in H, K and H-K. The top panels show the histograms for year 1 and the bottom panels for year 2.}
         \label{figure:5}
   \end{figure*}

The amplitude of the variations was measured as the peak-to-peak fluctuation in magnitudes and colour, for each year separately and also for the combined data. Figure~\ref{figure:5} shows histograms of the peak-to-peak magnitude amplitude for all the variable stars, members and candidate members. They show a peaked distribution, with most of the amplitude values ranging from 0.01 to 0.8 magnitudes. Table~\ref{table:5} shows the mean and maximum values of the amplitudes for the members and candidate members. When the variation is measured across the two years, the average amplitude for the variable members is of the order of a few tenths of a magnitude. One extreme object, AOC~J162636.81-241900.2, has been excluded from the calculations of Table~\ref{table:5} (see Sect.~4.2.5). The candidate member variables show mean \emph{K} and \emph{H}$-$\emph{K} amplitudes of the same order of the variable members within 1$\sigma$, and a mean \emph{H} amplitude within 2$\sigma$. There is no statistical evidence that members and candidate members represent different populations. \\

%__________________________________________________ Table of variability

\begin{table*}
\caption{Amplitudes of the Variability}           % title of Table
\label{table:5}      % is used to refer this table in the text
\centering                          % used for centering table
\begin{tabular}{ccccccccccc}        % centered columns 
\hline            % inserts double horizontal lines
\hline

~&~&\multicolumn{2}{c}{Year 1}&\multicolumn{2}{c}{Year 2}&\multicolumn{3}{c}{Years 1 and 2}\\
~&Band&Mean&Max.&Mean&Max.&Mean& $\sigma$$/$$\sqrt{n-1}$ &Max.\\
\hline   
Variable Candidates&\emph{H}&0.19&0.82&0.20&0.88&0.27&0.03&0.97\\
~&\emph{K}&0.15&0.62&0.15&0.56&0.21&0.02&0.62\\
~&\emph{H}$-$\emph{K}&0.09&0.36&0.08&0.36&0.15&0.02&0.52\\
\hline                                   %inserts single line						
Variable Candidates&\emph{H}&0.17&0.82&0.15&0.66&0.21&0.02&0.82\\
~&\emph{K}&0.13&0.61&0.13&0.61&0.18&0.02&0.65\\
~&\emph{H}$-$\emph{K}&0.14&0.97&0.12&0.70&0.18&0.02&0.97\\
\hline                        % inserts single horizontal line
\end{tabular}
\end{table*}
%______________________________________________   

\begin{figure}
   \centering
   \resizebox{\hsize}{!}{\includegraphics{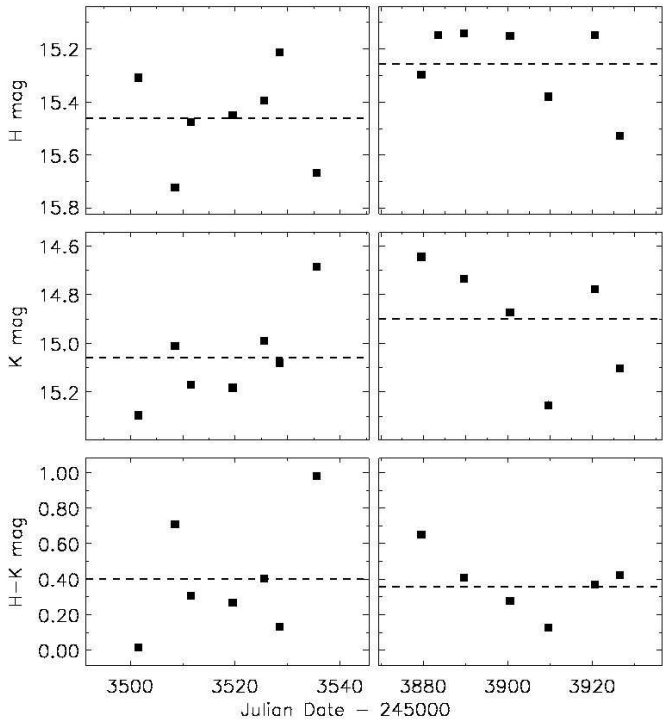}}
      \caption{AOC~J162814.73-242846.6; variable star with large variations in magnitude as well as in colour.}
         \label{figure:6}
   \end{figure}

\begin{figure}
   \centering
   \resizebox{\hsize}{!}{\includegraphics{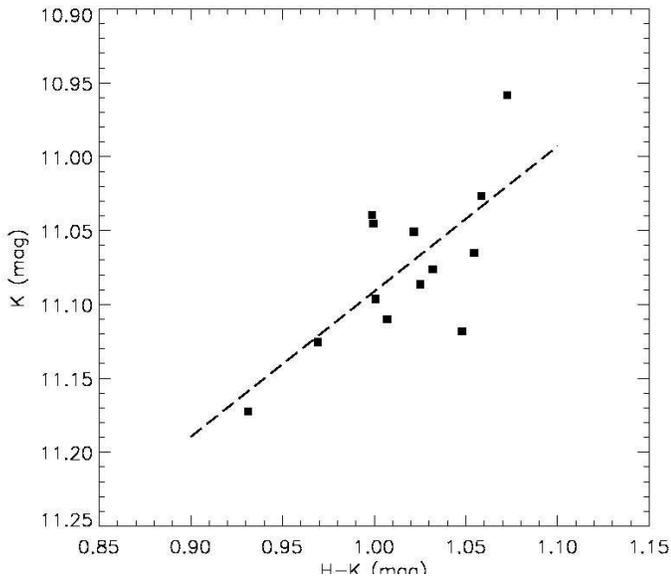}}
   \caption{AOC~J162812.72-241135.8; variable star with linearly correlated changes in \emph{H}$-$\emph{K} colour and brightness in \emph{K} band. Its colour gets bluer as the star gets fainter. The dashed line shows the linear fit.}
         \label{figure:7}
   \end{figure}

\begin{figure}
   \centering
   \resizebox{\hsize}{!}{\includegraphics{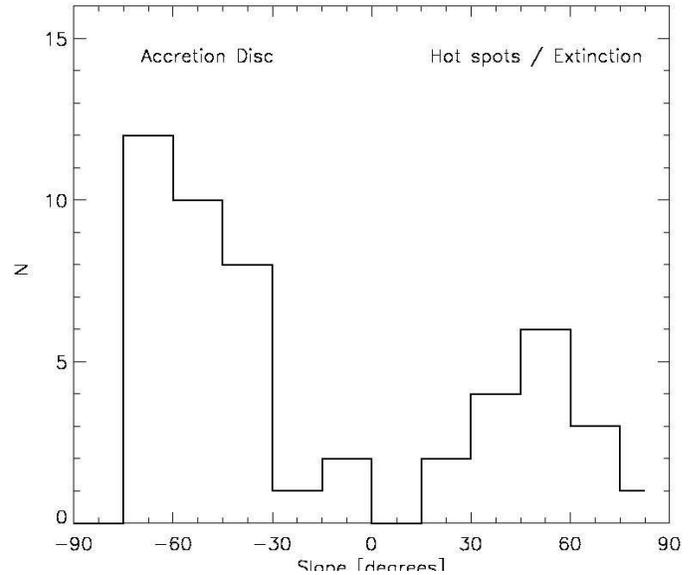}}
   \caption{Histogram of the derived slopes in the \emph{K} vs. \emph{H}$-$\emph{K} diagram. The slope can be used as a tool to distinguish the physical mechanisms behind the observed variations (Sect.~4.2). }
         \label{figure:8}
   \end{figure}

Most of the variable objects show a variation in \emph{H}$-$\emph{K} colour, which is on average 0.1 magnitudes but can be as large as 1 magnitude (see Fig.~\ref{figure:6}). In some cases, this variation is correlated with the change in brightness, and can be an important indicator of the underlying causes of variability \citep{Carpenter2001}. This behaviour is illustrated in Fig.~\ref{figure:7}, where the variations in the \emph{K}-band are linearly correlated with the changes in colour. The slopes in the colour-magnitude diagram \emph{K}~vs.~\emph{H}$-$\emph{K} were computed using the linear regression method for objects which showed a highly significant linear correlation between change in colour and brightness, i.e. with $<$~1\% probability that uncorrelated variables would yield a correlation coefficient at least as high as the assumed threshold. 36\% of all the variables, members and candidate members, satisfy the linear correlation criterion and the respective slopes are listed in Table~\ref{stars}. The angle convention used by \citet{Carpenter2001} was adopted, where a slope angle of 0$\degr$ represents a positive \emph{H}$-$\emph{K} colour change without a variation in \emph{K} magnitude, and increases clockwise. Figure~\ref{figure:8} is the histogram of the derived slopes which shows a binomial distribution with 16 objects having positive slopes, colour becoming redder as they fade, and 33 showing negative slopes, colour becomes bluer as the star gets fainter. These distinct behaviours can be associated with physical processes and will be discussed later in this paper (see Sect.~4.2). \\

%________________________________________________

%________________________________________________

%________________________________________________

\subsection{Infrared excesses}

Young stars show infrared emission which originates from dusty envelopes and circumstellar discs surrounding the central object. \citet{Lada1984}, based on the level of long wavelength excess with respect to a stellar photosphere emission, identified three different classes which define an IR-excess or SED classification scheme: Class~I, low mass protostars surrounded by an infalling envelope with large IR-excess; Class~II, young stars with accretion discs and a moderate IR-excess (as classical T~Tauri stars, CTTSs); and Class~III, stars which no longer accrete matter from a circumstellar disc and show no IR-excess (as weak-line T~Tauri stars, WTTSs). The IRAC data from Spitzer allows the study of these objects in the mid-IR, where the excess contribution from discs and envelopes is predominant. The IRAC colour-colour diagram ([3.6]$-$[4.5]~vs.~[5.8]$-$[8.0]) was presented as a tool to separate young stars of different classes \citep{Allen2004,Megeath2004}, and was already used in the study of other star forming regions as, for example, Taurus \citep{Hartmann2005,Luhman2006}, Serpens \citep{Harvey2007}, or Chamaeleon~II \citep{Porras2007}.

\begin{figure}
   \centering
   \resizebox{\hsize}{!}{\includegraphics{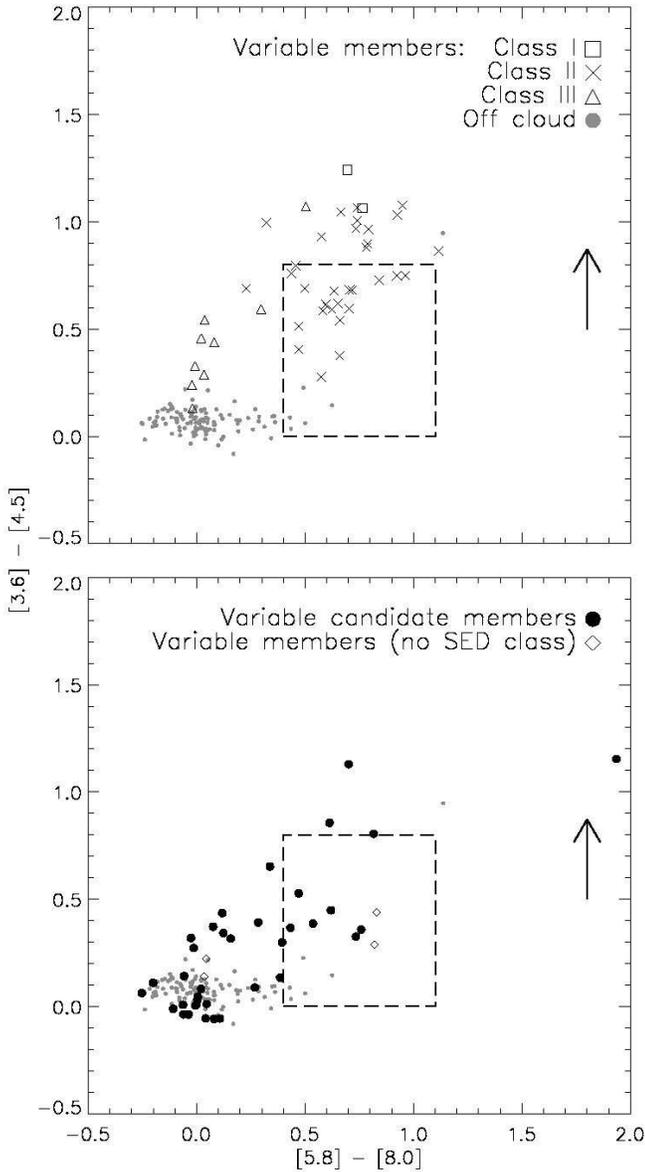}} 
      \caption{IRAC colour-colour diagram for variable members (top panel) and candidate members (bottom panel). Arrow represents the A$_{k}$=4 extinction vector for $\rho$~Ophiuchi \citep{Flaherty2007}.  }
         \label{figure:9}
   \end{figure}

The majority of population of $\rho$~Ophiuchi has been classified into SED classes. \citet{Bontemps2001} used ISOCAM mid-IR bands (6.7 and 15.3~$\mu$m) to detect and classify 212 sources. Using ground-base mid-IR observations, \citet{Barsony2005} confirmed those results. Also in the X-ray regime, Chandra and XMM-Newton observations \citep{Imanishi2001,Gagne2004,Ozawa2005} have contributed to the classification. The variability catalogue contains 48 variable members of $\rho$~Ophiuchi which were classified in these studies into Class~I, II, or III. Figure~\ref{figure:9} shows the IRAC colour-colour diagram for all variable objects with detections in the four bands, where the top diagram displays $\rho$~Ophiuchi members and the bottom diagram the candidate members. The variable members are displayed according to the SED class assigned in the literature, i.e., Class~I (open squares), Class~II (crosses), and Class~III (open triangles). The extinction vector is from \citet{Flaherty2007}. For comparison, a sample of objects has been chosen from a region of the sky away from the central cloud with little extinction which should mainly contain field stars.

The objects tend to cluster around three main regions of the diagram. Centred in the origin, [3.6]$-$[4.5],[5.8]$-$[8.0]=(0,0), are sources which have colours consistent with stellar photospheres and have no intrinsic IR-excess. These can be foreground and background stars, but also Class~III stars which do not have significant circumstellar dust. In this region of the colour-colour plane, it is not possible to differentiate between young stars and contaminants. However, as it can be seen in the top panel of the diagram, many of these objects are confirmed young members of $\rho$~Ophiuchi both from mid-IR and X-ray studies. Another preferred region for objects in the diagram is located within the box defined by \citet{Allen2004} which represents the colours expected from models of discs around young, low-mass stars. In fact, many of the Class~II objects classified by \citet{Bontemps2001} lie within that range. However, some sources previously identified as Class~II do not fall within the predicted limits. These objects have colours inconsistent with Class~II sources (higher [3.6]-[4.5] colours than Class~II but lower [5.8]$-$[8.0] colours than Class~I) and were previously classified as candidate flat spectrum objects. Their location in this diagram does not confirm them as transition objects between Class~I and II \citep{Bontemps2001} since they can be explained as reddened Class~II sources. Finally, from models of infalling envelopes,  \citet{Allen2004} predicts the colours of Class~I sources to have ([3.6]$-$[4.5])~$>$~0.8 and$/$or ([5.8]$-$[8.0])~$>$~1.1, which agrees well with the two Class~I objects identified with ISOCAM \citep{Bontemps2001}. Likewise, one of the sources classified with ISOCAM as Class~III lies on the Class~I$/$reddened Class~II region of the colour-colour diagram. In the ISOCAM paper, this source was classified as Class~III because it is located within the CS contours of the cloud, as defined by \citet{Liseau1995}. According to the IRAC diagram, its previous classification is incorrect. 

From the 84 candidate members, 37 are detected in the four IRAC bands and plotted in the bottom panel of Figure~\ref{figure:9} (filled circles). Four variable members are also detected in the four band but have no assigned SED class and therefore are also included (open diamonds). According to the criteria described above, they can be divided into Class~I (2 objects), Class~II (11 objects), and possibly Class~III (27 objects).

\begin{figure}
   \centering
   \resizebox{\hsize}{!}{\includegraphics{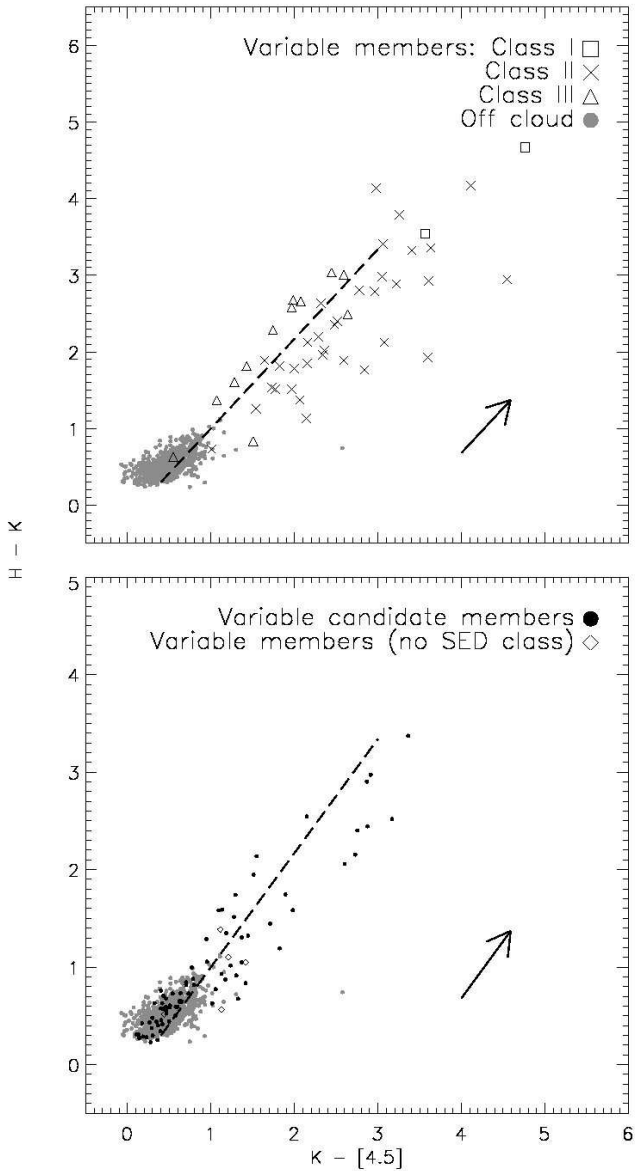}} 
      \caption{WFCAM/IRAC colour-colour diagram. Arrow represents the A$_{V}$=10 extinction vector for $\rho$~Ophiuchi \citep{Flaherty2007}.  }
         \label{figure:10}
   \end{figure}

The number of variables with detections in the 4.5~$\mu$m IRAC/Spitzer band is higher than for longer wavelengths, since this band has a higher sensitivity. Figure~\ref{figure:10} shows the \emph{H}$-$\emph{K} vs. \emph{K}$-$[4.5] colour-colour diagram for all variable objects with detections in the IRAC/Spitzer 4.5~$\mu$m band, where again the top diagram displays the members of $\rho$~Ophiuchi and the bottom diagram the candidate members. The variable members are displayed according to their SED class from the literature (same convention as in Fig.~\ref{figure:9}). The reddening vector is from \citet{Flaherty2007} and the dash line follows the same reddening law. The top panel of Fig.~\ref{figure:10} shows a clean break between class I/II and class III stars (i.e., stars with and without circumstellar dust emission) for \emph{K}$-$[4.5]$>$1 and \emph{H}$-$\emph{K}$>$0.5. Applying this criterion to the variable candidate members (bottom panel), 23 candidate members can be classified as ClassI/II, and 8 as possible class III. Furthermore, for objects present in both Fig.~\ref{figure:9} and Fig.~\ref{figure:10}, the SED classes determined are in agreement. The only exception are 3 variable candidate members which show a significant IR-excess in the \emph{K}$-$[4.5] colour but less at longer wavelengths.

Combining WFCAM near-IR observations with IRAC/Spitzer mid-IR data, it is possible to classify 60\% of the candidate members as Class I/II and possible Class III. Furthermore, the SED classes from the literature for the variable members are confirmed, with the exception of few objects (see discussion above).

\section{Discussion}

\subsection{Membership}

\begin{figure}
   \centering
   \resizebox{\hsize}{!}{\includegraphics{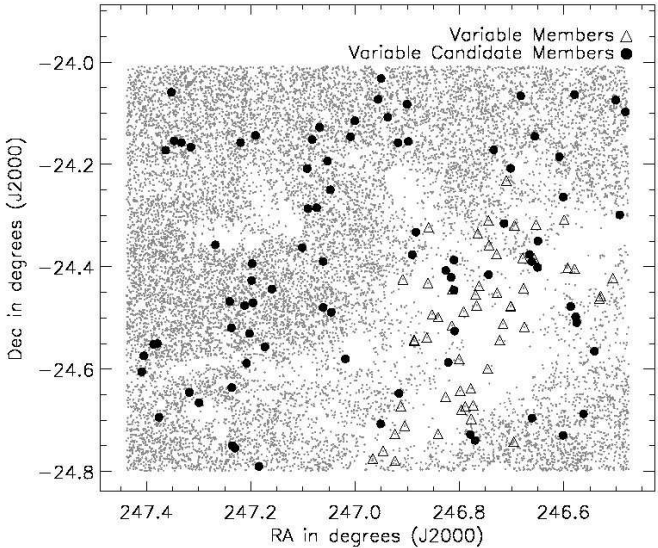}}
      \caption{Spatial distribution for variable members of $\rho$~Ophiuchi and variable candidates, with respect to all detected objects in the field. }
         \label{figure:11}
   \end{figure}

The nature  of the candidate member population is very important, since it determines the reliability of variability as an indicator of youth. In principle, some or all of the candidate members could be background objects -- either variable stars within our own galaxy or extragalactic variable objects, principally Active Galactic Nuclei (AGN).  An inspection of Fig.~\ref{figure:11}, however, shows that this is unlikely to be the case.  The candidate members do not appear to be distributed in the same way as the majority of background point sources against which the clouds can be seen -- they are noticeably absent from the regions of lowest extinction for example, and appear to be more clustered. To put this on a quantitative basis the 2-D Kolmogorov-Smirnov (K-S)  test as outlined by \citet{Peacock1983} was applied to candidate member and full point-source detection (non-variable) samples -- the latter clearly dominated by the background population of stars and galaxies. The test is approximately distribution-free, which means the probability that the two samples are from the same population can be estimated.  For 84 candidate member objects, a Z statistic of 1.99 was determined against the background distribution. \citet{Peacock1983} gives a formula for calculating the significance level, which shows that there is an 2\% chance that the two are from the same population. This is an upper limit since, for reasons already discussed, the variable candidate members lie outside the regions of high extinction, as do the background sources. This introduces an unavoidable spatial bias, making the two distributions look similar. The K-S test suggests the two populations come from different distributions. Of course this is for the candidate population as a whole -- there is still the possibility that a subset are due to contamination from background variable objects.

AGN are a potentially important  source of contamination given the faint magnitude limits of the sample. The \emph{K}~$=$~18.0 limit amounts to 0.08~mJy at 3.6~$\mu$m assuming an SED median index of $-$1.2 \citep{Buchanan2006}. \citet{Treister2006} use a Spitzer survey of the GOODS field to estimate AGN number density versus flux in the IRAC bands. Their data show that this flux limit should result in approximately 300 AGN detections per sq. degree. However, central to understanding contamination by AGN is the distribution and magnitude of extinction in $\rho$~Ophiuchi. \citet{Kenyon1998}, in a study of the reddening law, find that in the clear areas well outside the molecular gas contours in areas of least extinction, the \emph{H}$-$\emph{K} colour in most sources has only a small excess over intrinsic values -- of order 0.2 in \emph{H}$-$\emph{K}. A similar excess is found in these data, corresponding to a visual extinction of 3-4 magnitudes. However, over much of the field the extinction is much higher. Therefore, for a proper evaluation, the number counts of \emph{K}-band point sources (Fig.~\ref{figure:1}) was used to calculate \emph{K} band extinction in a grid of 10~x~10 squares over the observed field. This was then used with the AGN cumulative flux distribution \citep{Treister2006} to give the integrated number of AGN expected over the WFCAM field of view. After integrating over the extinction distribution a total of 33 AGN should be present in the $\rho$~Ophiuchi field to the \emph{K} band detection limit of 18.0. How many of these will be detected as variable? With the steep AGN cumulative flux distribution, more than 95\% of the AGNs will lie in the faintest two magnitude bins, where the typical photometric error of the sample is 0.05-0.1 magnitudes. Simulations with the 2-band cross-correlation index show  that a clear variability detection with this index requires an amplitude of at least $>$~0.13 magnitudes in this case. \citet{Enya2002} show that the fraction of AGN with long-term IR variability larger than this is ~28\% or 9 objects in the $\rho$~Ophiuchi field, compared to the candidate-member sample of 84. So while a precise number is difficult to calculate, it is  likely that a small (but still non-negligible) fraction of the variable candidate-members can be explained as AGN.

Another potential source of contamination is a population of background variable stars. The galactic model of \citet{Wainscoat1992} was used as a useful tool in investigating this. Even though discrepancies between observations and this model exist, it is certainly accurate enough (factor of 2) for the purposes here. \citet{Schultheis2000} conducted an unbiased survey of a region in the galactic bulge, and found 720 sources variable in \emph{J} and \emph{K} band, representing 0.2\% of the source detections in the region. These were mainly long-period variables -- AGB stars at or beyond the tip of the RGB. A similar fraction of variables in the $\rho$~Ophiuchi field (15\,191 sources) would predict 24 variables. However, the stellar populations are very different in the two fields. The bulge field is dominated by giants and includes many AGB stars; in fact the galactic model predicts a number of AGB stars comparable to the total number of variables detected in the \citet{Schultheis2000} study. But the high galactic latitude of the $\rho$~Ophiuchi field (+16.7) completely changes the result. Far fewer giants are expected, since the line of sight moves rapidly above the plane and bulge, so that the counts become dominated by the faint end of the dwarf luminosity function. At this latitude, the model predicts $<$~1 AGB star/long-period variable in the whole field of view. It seems clear that the $\rho$~Ophiuchi variable sample should be completely uncontaminated by these objects. 

It should be noted that in discussing the membership of any individual source, its location in the field is important, since any significant amount of extinction lowers the probability of background contamination substantially compared to the numbers calculated above. \\

\subsection{Understanding the Variability}

The characteristics of the variability observed in young stars in $\rho$~Ophiuchi are very diverse, spanning large ranges in magnitude, colour and timescale. Furthermore, given the limited information and finite measurement errors, it is not possible to conclude with certainty which physical mechanism is causing variability in each star. In fact, in many cases the variability observed in a particular star could be caused by a variety of mechanisms. So the following approach was used. Known or plausible physical mechanisms were considered in turn as the cause of variability in each star. Mechanisms which were considered plausible according to certain criteria were then listed in Table~\ref{stars} for each star. In this way, an overall idea of whether the variability can be explained by conventional mechanisms, is obtained. A summary of the mechanisms, criteria and conclusions is as follows.

\subsubsection{Rotational modulation by cool starspots}

Cool starspots are known to exist both in CTTSs and WTTSs, and to cause variations in the stellar brightness. They are colder than the photosphere and arise from magnetic active regions, analogously to solar sunspots \citep{Bouvier1989}. The fractional area coverage is of the order of 20\% \citep{Bouvier1993,Herbst1994} which determines the maximum amplitude of 0.4 magnitudes \citep{Carpenter2001,Herbst2002}. \citet{Herbst1994} presented a classification scheme for optical variability, defining as type~I variable stars with low, periodic, amplitude magnitude variations (a few tenths of a magnitude) caused by cool spots. Photometric variability studies of very low mass stars and brown dwarfs have identified similar behaviours \citep[e.g.,][]{Scholz2005}. Using single-temperature blackbody models and input parameters inferred from optical studies to model stellar spots behaviours, \citet{Carpenter2001} find that the expected amplitude colour variations produced by cool spots do not exceed 0.05 magnitudes both in \emph{J}$-$\emph{H} and \emph{H}$-$\emph{K} colours. 

In the $\rho$~Ophiuchi variable population, stars with amplitudes in \emph{H} and \emph{K}~$<$~0.4 magnitudes and \emph{H}$-$\emph{K} colour variations $<$~0.05 magnitudes show variations consistent with the existence of cool spots, as defined by the above parameters, and are classified as CS (cool spots; see Col.~7, Table~\ref{stars}). These amount to 17\% of the variable members and 21\% of the candidate members. Late-type field stars with magnetic activity could, however, still contaminate this group of variables (Sect.~3.1). 

\subsubsection{Rotational modulation by hot starspots}

Hot spots are interpreted as the impact points on the stellar surface from disc accretion through magnetic field lines \citep[e.g.,][]{Calvet1992}. They cover a smaller fraction of the stellar surface but the high temperatures can cause larger amplitude variations, which in the near-IR regime can be as high as 0.2-0.4 magnitudes for the \emph{JHK} bands, and between ~0.05-0.12 in colours \citep{Carpenter2001}. The timescales are shorter compared to cool spots \citep[e.g.,][]{Kenyon1994}. In the \citet{Herbst1994} classification scheme, a type~II variability class is assigned to objects showing larger amplitude variations (which can be irregular or periodic), from short-lived hot spots. To analyse the behaviour produced by hot spots, \citet{Carpenter2001}) use the information from the correlated colour and magnitude changes.  The predicted slope from hot spot models has high, positive values, although given the simplicity of their models, it is not possible to identify a specific range.

Variable stars with amplitudes in \emph{H} and \emph{K}~$<$~0.4 magnitudes, \emph{H} and \emph{K} colour variations ~0.05-0.12 magnitudes, and a positive slope in the colour-magnitude diagram (if a slope was determined) are classified as HS (hot spots). Hot spots can explain variability in 28\% of the variable members and 30\% of the candidate members.

\subsubsection{Circumstellar Extinction}

Variations in extinction can originate from the intersection with the line-of-sight of infalling or orbiting material in the circumstellar environment \citep{Skrutskie1996}. Circumstellar extinction variations will follow a reddening law if the grain size of the material is comparable to the interstellar grain size distribution \citep{Skrutskie1996}. The interstellar reddening law E(\emph{H}$-$\emph{K})/\emph{A}$_K$~=~0.56 was adopted from \citet{Rieke1985}. In the \emph{K} vs. \emph{H}$-$\emph{K} diagram, this means that extinction variations can explain photometric slopes of ~60\degr, with the object being redder when it is faint. The histogram of the colour-magnitude slopes (Fig.~\ref{figure:6}) peaks at 45\degr-60\degr, showing that variations in extinction can explain most of the observed positive slopes. 

Variable stars with positive slopes between 45\degr and 60\degr are classified as E (extinction). These represent 6\% of the variable members and 4\% of the variable candidate members.

\subsubsection{Accretion Discs}

Changes in the disc structure caused by mechanisms such as thermal instabilities, variable magnetic fields or warping instability can lead to near-IR variations on short timescales \citep[see][and references therein]{Carpenter2001}. Using simple disc models to study the consequences expected from geometric variations in a circumstellar disc, \citet{Carpenter2001}  find that the observed negative slope in the colour-magnitude correlation (Fig.~\ref{figure:5}) can be explained by changes in the inner disc structure.  This scenario has also been suggested by \citet{Eiroa2002}. 

Variable stars for which a negative colour-magnitude slope was found, where the star is bluer when faint, are classified as AD (accretion disc variations). In  $\rho$~Ophiuchi, 23\% of the variable members and 25\% of the variable candidate members have a negative slope.

\subsubsection{Extreme variability}

Although very rarely observed, another possible cause of variability related to accretion discs, is an Fu Orionis type outburst \citep[e.g.][and references therein]{Reipurth2007}. AOC~J162636.81-241900.2 (Fig.~\ref{figure:11}) has been previously classified as a member of $\rho$~Ophiuchi. Its light curve reveals a brightening of several magnitudes (3.5 magnitudes in the \emph{K}-band) in approximately one year, in what could be a Fu Orionis type burst. However, a minimum in \emph{H} and \emph{K} magnitudes of 13.93 and 12.31, respectively, have previously been reported by \citet{Barsony1998}, meaning this object could instead be an EXor, which are stars that undergo a maximum in magnitude once every few years, thought to be caused by the massive infall of circumstellar material onto the central star \citep[see][and references therein]{Herbig2008}. A spectroscopy study of this variable member (classified as EV) is needed to better understand its nature.

\begin{figure}
   \centering
\     \resizebox{\hsize}{!}{\includegraphics{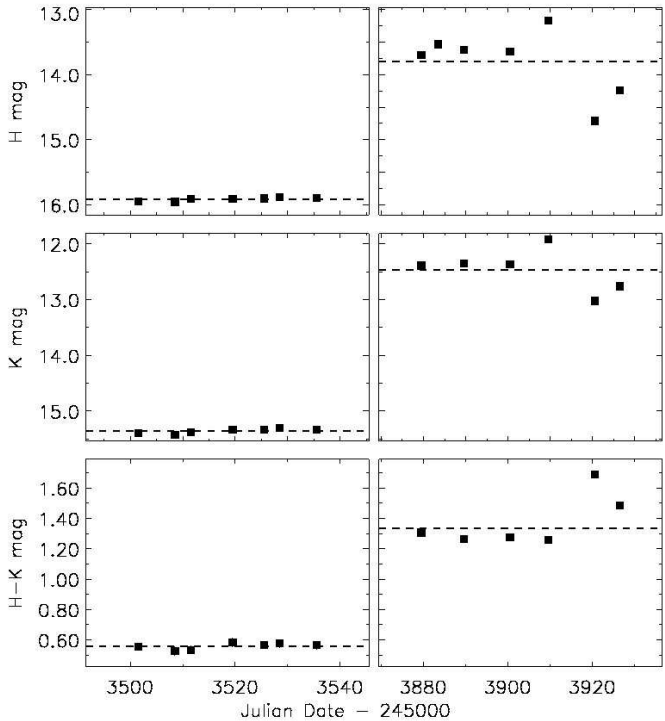}}
      \caption{AOC~J162636.81-241900.2: variable member with a brightening of several magnitudes in both \emph{H} and \emph{K} bands.}
         \label{figure:12}
   \end{figure}

\subsubsection{Unexplained variability}

There are 32\% of the members and 26\% of the candidate members which do not fit into any of the criteria defined above. Among the members, 15 objects out of 17 have colours consistent with SED Class I/II stars. However the majority does not show a linear correlation in the colour-magnitude variability diagram. The most likely explanation is that several physical processes are taking place simultaneously, and the variability observed is the combination of all the intervenients.  This scenario is further supported by the fact that 4 of these variable members show a positive slope which is either too shallow or to steep to be explained by changes in extinction, and colours inconsistent with hot spots rotational modulation. The same is true for the only candidate member with a determined, positive slope, which has also been classified as Class~II, but does not fit into the variability categories. All the other candidate members with unexplained variability did not show a linear correlation in the colour-magnitude variability diagram, and very few have IRAC/Spitzer detections. The other characteristic which is valid for these objects, both members and candidate members, is that the mean \emph{H} and \emph{K} amplitude magnitudes, as well as the \emph{H}$-$\emph{K} colour variations, are twice as large as the ones seen among other variables. Although this is in conformity with the scenario just described, it also opens the possibility that more extreme variability behaviours are, after all, common among young stars. Only targeted photometric and spectroscopic monitoring of variable stars of different types, looking for a correlation between colours, magnitudes, and spectral features would help in isolating the important physical mechanisms.

%_________________________________________________________________

\section{Conclusions}

A multi-epoch, very deep near-IR survey was conducted in the Ophiuchus molecular cloud with the WFCAM/UKIRT. Statistical methods, such as the reduced chi-square and correlation indices, were used in the search for variability, known to be a defining characteristic of young stellar objects. 137 variable objects were found which show timescales of variation which can go from days to years, amplitude magnitude changes from a few tenths to ~3 magnitudes, and significant colour variations. From the $\rho$~Ophiuchi known population, 128 members have counter-parts in the WFCAM catalogue, and 41\% are found to be variable. 

The dataset was merged with IRAC/Spitzer observations to further extend the information about the detected variables and using colour-colour diagrams, ([3.6]$-$[4.5]~vs.~[5.8]$-$[8.0]) and/or (\emph{H}$-$\emph{K}~vs.~\emph{K}$-$[4.5]), 50 variable candidate members were classified as candidate pre-main-sequence stars (23 Class~I/II and 27 possible Class~III variables).
 
The variability trends found were interpreted in a physical context by comparison with previous suggested classifications for variability in the optical and near-IR. The characteristics observed are consistent with the existence of cool or/and hot spots on the stellar surface, variations in circumstellar extinction, or structural variations in accretion discs. However, a large fraction of the variable population does not fit into the predicted parameters for near-IR variability, which can be explained by the fact that the variability observed does not reflect a single event, but the net effect of several simultaneous processes.

Finally, near-IR variability was used to discover a candidate population of pre-main-sequence stars associated with the $\rho$~Ophiuchi star forming region. The colour-magnitude diagram, \emph{K} vs. \emph{H}-\emph{K}, provides evidence that many of the newly found low-luminosity variables have colours consistent with those expected from young, very low mass objects. To confirm the nature of these variable objects, a low resolution near-IR spectroscopic follow-up has been conducted for a small sample and results will be present in an accompanying paper.

%__________________________________________________________________

\begin{acknowledgements}
We thank the UKIRT observatory staff and WFCAM Science Archive team for acquiring and pipeline processing the observations used for this project. C.~A.~O. acknowledges partial support from a Marie Curie Fellowship for Early Research Training. This work is based in part on observations made with the Spitzer Space Telescope, which is operated by the Jet Propulsion Laboratory, California Institute of Technology under a contract with NASA.
\end{acknowledgements}

\bibliographystyle{aa}
\bibliography{cat}

%__________________________________________________ Table of variability

% If table 2
\longtab{4}{

%\begin{landscape}     
\begin{longtable}{c c c c c c c c c c c}
\caption{\label{stars} Photometric Properties of the Variable Stars}\\
\hline\hline
AOC designation\footnotemark[1] & $\textit{$<$H$>$}$\footnotemark[2]     & $\textit{$<$K$>$}$\footnotemark[2]   &  $\Delta\textit{H}$\footnotemark[3]  &  $\Delta\textit{K}$\footnotemark[3] & $\Delta(\textit{H}$-$\textit{K})$\footnotemark[3] & Slope\footnotemark[4] & SED\footnotemark[5] & \multicolumn{2}{c}{Variability\footnotemark[6]\footnotemark[7]} & Ref.\footnotemark[8] \\
~ & (mag)          & (mag)        & (mag)          & (mag)          & (mag)        & (degrees) &  ~ &  ~ &  ~ &  ~ \\

\hline
\endfirsthead
\caption{continued.}\\
\hline\hline
AOC designation\footnotemark[1] & $\textit{$<$H$>$}$\footnotemark[2]     & $\textit{$<$K$>$}$\footnotemark[2]   &  $\Delta\textit{H}$\footnotemark[3]  &  $\Delta\textit{K}$\footnotemark[3] & $\Delta(\textit{H}$-$\textit{K})$\footnotemark[3] & Slope\footnotemark[4] & SED\footnotemark[5] & \multicolumn{2}{c}{Variability\footnotemark[6]\footnotemark[7]} & Ref.\footnotemark[8] \\
~ & (mag)          & (mag)        & (mag)          & (mag)          & (mag)        & (degrees) &  ~ &  ~ &  ~ &  ~ \\

\hline
\endhead
\hline
\endfoot

J162555.67-240549.6 & 13.97 & 13.47 & 0.05 & 0.06 & 0.04 & ... & ...  & Y00Y & CS    & ...     \\
J162558.29-241754.9 & 17.23 & 15.16 & 0.45 & 0.26 & 0.50 & ... & ...  & YYYY & ...   & ...     \\
J162600.16-240427.3 & 14.15 & 13.71 & 0.09 & 0.09 & 0.03 & ... & ...  & Y0YY & CS    & ...     \\
J162601.37-242520.4 & 12.30 & 11.04 & 0.45 & 0.29 & 0.17 & 54  & II   & YYYY & E     & 2       \\
J162607.03-242724.1 & 12.42 & 10.40 & 0.44 & 0.31 & 0.15 & 65  & II   & YYYY & ...   & 2,3     \\
J162607.63-242741.4 & 11.88 & 10.27 & 0.17 & 0.16 & 0.06 & ... & III  & YYYY & HS    & 2,3     \\
J162609.92-243354.2 & 11.77 & 10.89 & 0.21 & 0.05 & 0.17 & ... & II   & YYY0 & ...   & ...     \\
J162615.01-244114.8 & 15.19 & 14.31 & 0.31 & 0.26 & 0.22 & ... & ...  & YYYY & ...   & ...     \\
J162618.13-243033.1 & 12.98 & 11.67 & 0.06 & 0.08 & 0.04 & -53 & II   & Y0Y0 & CS,AD & ...     \\
J162618.57-242951.4 & 15.13 & 13.55 & 0.07 & 0.08 & 0.07 & ... & II   & Y0Y0 & HS    & ...     \\
J162618.98-242414.0 & 14.28 & 12.15 & 0.54 & 0.47 & 0.18 & ... & II   & YYYY & ...   & 1,2     \\
J162619.10-240348.4 & 11.42 & 10.94 & 0.11 & 0.05 & 0.12 & ... & III  & Y00Y & HS    & ...     \\
J162620.83-242839.5 & 13.61 & 11.87 & 0.63 & 0.47 & 0.15 & 59  & II   & YYYY & E     & ...     \\
J162622.26-242407.0 & 14.83 & 13.69 & 0.10 & 0.14 & 0.05 & -69 & II   & YYY0 & CS,AD & 1,2     \\
J162623.81-241829.0 & 14.59 & 13.49 & 0.12 & 0.04 & 0.08 & 20  & II   & Y0Y0 & HS    & 1       \\
J162624.29-241549.7 & 15.81 & 13.65 & 0.14 & 0.16 & 0.03 & ... & II   & YYYY & CS    & ...     \\
J162624.41-244345.5 & 16.44 & 15.71 & 0.23 & 0.32 & 0.37 & -36 & ...  & YYY0 & AD    & ...     \\
J162626.08-241106.3 & 15.20 & 14.54 & 0.20 & 0.20 & 0.11 & ... & ...  & YYYY & HS    & ...     \\
J162635.95-242058.7 & 15.99 & 13.59 & 0.51 & 0.53 & 0.31 & ... & II   & YYYY & ...   & ...     \\
J162636.07-242404.2 & 16.08 & 13.18 & 0.16 & 0.13 & 0.09 & ... & II   & Y0YY & HS    & ...     \\
J162636.81-241900.2 & 14.86 & 14.02 & 2.78 & 3.52 & 1.16 & -72 & II   & YYYY & AD,EV & 2,3     \\
J162637.41-240842.1 & 11.21 & 10.46 & 0.14 & 0.04 & 0.13 & ... & III  & YYYY & ...   & ...     \\
J162637.78-242300.7 & 12.60 & 10.72 & 0.77 & 0.44 & 0.34 & ... & II   & YYYY & ...   & 2       \\
J162638.66-244144.4 & 14.96 & 14.15 & 0.09 & 0.10 & 0.10 & ... & ...  & Y00Y & HS    & ...     \\
J162638.79-242322.8 & 12.84 & 11.52 & 0.06 & 0.07 & 0.04 & ... & II   & Y00Y & CS    & ...     \\
J162639.92-242233.5 & 14.69 & 13.49 & 0.08 & 0.04 & 0.06 & ... & II   & Y00Y & HS    & ...     \\
J162642.14-243103.0 & 13.68 & 11.05 & 0.88 & 0.56 & 0.36 & ... & II   & YYYY & ...   & 2       \\
J162642.51-242631.7 & 15.63 & 12.60 & 0.06 & 0.08 & 0.05 & ... & III  & Y0Y0 & HS    & 1,2,3   \\
J162642.89-242259.1 & 12.91 & 11.37 & 0.04 & 0.06 & 0.03 & ... & II   & Y0Y0 & CS    & 2,3     \\
J162643.90-240357.8 & 16.28 & 15.69 & 0.19 & 0.16 & 0.33 & -24 & ...  & Y00Y & AD    & ...     \\
J162646.76-241908.4 & 11.68 & 10.29 & 0.06 & 0.07 & 0.05 & ... & III  & Y0Y0 & CS    & 3       \\
J162647.05-244430.0 & 11.20 & 10.56 & 0.10 & 0.03 & 0.11 & ... & III  & Y0YY & HS    & 2,4,5   \\
J162648.39-242834.8 & 15.24 & 12.46 & 0.11 & 0.09 & 0.07 & ... & II   & Y00Y & HS    & 2,3     \\
J162648.48-242838.9 & 14.07 & 11.09 & 0.49 & 0.31 & 0.21 & ... & II   & YYYY & ...   & 1,2,3   \\
J162648.53-241227.5 & 11.84 & 10.55 & 0.10 & 0.08 & 0.05 & ... & III  & Y0YY & HS    & ...     \\
J162650.47-241352.3 & 11.32 & 10.81 & 0.10 & 0.04 & 0.11 & ... & III  & Y00Y & HS    & 5       \\
J162651.52-241855.1 & 17.39 & 15.80 & 0.52 & 0.28 & 0.47 & ... & III  & Y0Y0 & ...   & ...     \\
J162651.97-243039.6 & 16.55 & 13.23 & 0.08 & 0.11 & 0.06 & -57 & II   & Y0YY & AD    & 1,2,3   \\
J162653.48-243236.2 & 16.93 & 13.15 & 0.20 & 0.20 & 0.16 & ... & III  & YY0Y & ...   & 1,2,3   \\
J162654.77-242702.3 & 14.97 & 12.77 & 0.21 & 0.18 & 0.11 & ... & II   & YYYY & HS    & 2       \\
J162654.98-242229.7 & 11.92 & 10.11 & 0.07 & 0.06 & 0.06 & ... & III  & Y0YY & HS    & 2,3     \\
J162656.27-241017.6 & 14.91 & 14.09 & 0.12 & 0.12 & 0.04 & ... & ...  & Y0YY & CS    & ...     \\
J162658.39-242130.2 & 13.18 & 11.37 & 0.18 & 0.20 & 0.08 & ... & II   & YYYY & HS    & 2,3     \\
J162658.64-241834.8 & 13.13 & 11.61 & 0.05 & 0.08 & 0.04 & -63 & II   & Y0Y0 & CS,AD & 2,3     \\
J162658.65-242455.5 & 17.02 & 14.50 & 0.22 & 0.22 & 0.08 & ... & II   & YYYY & HS    & ...     \\
J162659.04-243556.9 & 13.87 & 11.99 & 0.38 & 0.19 & 0.20 & 42  & II/III& YYYY & ...   & 1,2     \\
J162702.99-242614.7 & 15.83 & 12.42 & 0.25 & 0.20 & 0.11 & ... & II   & YYYY & HS    & 2,3     \\
J162703.58-242005.5 & 14.88 & 13.51 & 0.34 & 0.29 & 0.21 & ... & II   & YYYY & ...   & 2       \\
J162704.10-242829.9 & 13.12 & 10.77 & 0.19 & 0.16 & 0.15 & ... & II   & YYYY & ...   & 1,2,3   \\
J162704.57-242715.6 & 13.22 & 11.10 & 0.39 & 0.28 & 0.17 & 47  & II   & YYYY & E     & 2,3     \\
J162704.88-244421.2 & 15.72 & 14.90 & 0.11 & 0.08 & 0.09 & ... & ...  & Y0YY & HS    & ...     \\
J162705.66-244013.1 & 16.81 & 13.81 & 0.14 & 0.10 & 0.09 & ... & III  & Y0Y0 & HS    & 2       \\
J162706.59-244148.9 & 11.43 & 10.71 & 0.21 & 0.11 & 0.22 & ... & II   & YYYY & ...   & 2,4,5   \\
J162706.76-243815.1 & 14.19 & 11.25 & 0.38 & 0.46 & 0.13 & -70 & II   & YYYY & AD    & 1,2,4   \\
J162707.02-244338.9 & 15.61 & 14.61 & 0.03 & 0.04 & 0.05 & -39 & ...  & 0Y0Y & AD    & ...     \\
J162709.34-244022.5 & 13.72 & 11.32 & 0.08 & 0.14 & 0.07 & -61 & II   & YYYY & AD    & 2       \\
J162710.02-242913.3 & 15.21 & 14.16 & 0.29 & 0.21 & 0.10 & ... & ...  & YYYY & HS    & 1       \\
J162711.16-244046.7 & 13.47 & 10.55 & 0.97 & 0.62 & 0.52 & ... & II   & YYYY & ...   & 2       \\
J162711.71-243832.1 & 15.10 & 10.93 & 0.77 & 0.49 & 0.36 & 40  & II   & YYYY & ...   & 1,2,4   \\
J162712.12-243449.1 & 13.09 & 11.32 & 0.79 & 0.55 & 0.27 & 58  & II   & YYYY & E     & 2,4     \\
J162714.34-243131.9 & 16.65 & 15.20 & 0.41 & 0.30 & 0.15 & 57  & II   & Y0YY & E     & ...     \\
J162714.49-242646.0 & 15.46 & 12.08 & 0.17 & 0.25 & 0.19 & -55 & I/II & YYY0 & AD    & ...     \\
J162714.61-242312.6 & 16.65 & 15.14 & 0.12 & 0.22 & 0.15 & -51 & III  & YYY0 & AD    & ...     \\
J162715.44-242639.8 & 13.43 & 10.55 & 0.33 & 0.25 & 0.24 & ... & II   & YYYY & ...   & 4       \\
J162715.50-243053.8 & 16.49 & 12.35 & 0.19 & 0.28 & 0.16 & -53 & II/III& YYYY & AD    & 2       \\
J162715.87-242514.1 & 16.12 & 13.15 & 0.33 & 0.27 & 0.11 & 61  & II   & YYYY & HS    & ...     \\
J162717.19-243512.8 & 16.59 & 14.65 & 0.14 & 0.03 & 0.15 & ... & III  & Y00Y & ...   & ...     \\
J162718.36-242426.3 & 13.34 & 11.20 & 0.11 & 0.08 & 0.07 & ... & III  & Y0YY & HS    & ...     \\
J162718.37-243914.8 & 15.60 & 12.05 & 0.17 & 0.15 & 0.07 & ... & II   & YYYY & HS    & 1,2,4   \\
J162721.78-242953.4 & 15.18 & 10.51 & 0.64 & 0.43 & 0.41 & 40  & I/II & YYYY & ...   & 1,2,4   \\
J162721.81-244335.8 & 13.40 & 10.74 & 0.06 & 0.06 & 0.03 & ... & III  & Y0YY & CS    & 2,4     \\
J162724.63-242935.5 & 14.99 & 12.40 & 0.12 & 0.11 & 0.05 & ... & III  & YYYY & CS    & 1,4     \\
J162726.22-241923.1 & 14.42 & 12.91 & 0.08 & 0.09 & 0.08 & -46 & II   & Y0YY & AD    & 2       \\
J162726.57-242554.4 & 12.33 & 11.77 & 0.18 & 0.23 & 0.10 & -66 & II   & YYYY & AD    & 5       \\
J162727.06-243217.6 & 14.59 & 12.30 & 0.07 & 0.07 & 0.05 & -47 & III  & Y0Y0 & CS,AD & 4       \\
J162732.05-241956.3 & 17.21 & 15.47 & 0.18 & 0.06 & 0.17 & ... & III  & Y00Y & ...   & ...     \\
J162732.70-243242.4 & 15.49 & 13.00 & 0.51 & 0.37 & 0.28 & ... & II   & YYYY & ...   & 2       \\
J162732.84-243234.9 & 12.86 & 10.90 & 0.12 & 0.12 & 0.09 & ... & II   & YYYY & HS    & 2,4     \\
J162733.75-242234.9 & 16.49 & 15.66 & 0.60 & 0.38 & 0.28 & 52  & II   & YY0Y & E     & ...     \\
J162735.58-240918.7 & 14.85 & 14.11 & 0.05 & 0.07 & 0.04 & -62 & ...  & Y0Y0 & CS,AD & ...     \\
J162736.14-240457.7 & 12.97 & 12.06 & 0.28 & 0.21 & 0.09 & ... & II   & YYYY & HS    & ...     \\
J162737.23-244237.9 & 14.77 & 11.41 & 0.18 & 0.27 & 0.17 & -58 & II   & YYYY & AD    & 2       \\
J162738.06-242527.8 & 15.37 & 12.70 & 0.12 & 0.04 & 0.09 & 20  & III  & Y00Y & HS    & 2       \\
J162738.95-244020.7 & 14.19 & 12.42 & 0.23 & 0.35 & 0.14 & -63 & II   & YYYY & AD    & 2,4     \\
J162739.88-243851.3 & 15.41 & 14.06 & 0.05 & 0.06 & 0.04 & ... & III  & Y0Y0 & CS    & ...     \\
J162740.28-240927.2 & 14.47 & 13.63 & 0.04 & 0.06 & 0.04 & -56 & III  & Y0Y0 & CS,AD & ...     \\
J162741.60-244644.8 & 15.78 & 13.84 & 0.39 & 0.15 & 0.32 & ... & II   & YYYY & ...   & 2       \\
J162741.74-244336.2 & 14.69 & 11.88 & 0.31 & 0.22 & 0.12 & ... & II   & YYYY & ...   & 2       \\
J162744.98-240628.6 & 11.78 & 10.74 & 0.66 & 0.44 & 0.47 & ... & II/III& YYYY & ...   & ...     \\
J162747.08-244535.1 & 12.94 & 11.09 & 0.08 & 0.08 & 0.04 & ... & II   & Y0YY & CS    & 2,4     \\
J162748.11-240155.9 & 16.49 & 15.90 & 0.48 & 0.38 & 0.72 & ... & ...  & 0YYY & ...   & ...     \\
J162748.24-244225.6 & 16.92 & 14.86 & 0.54 & 0.52 & 0.18 & ... & I/II & YYYY & ...   & ...     \\
J162749.45-240422.2 & 14.19 & 13.61 & 0.07 & 0.05 & 0.04 & ... & III  & Y00Y & CS    & ...     \\
J162751.90-244629.7 & 11.82 & 10.45 & 0.06 & 0.06 & 0.04 & ... & III  & Y0YY & CS    & 2,4     \\
J162800.20-240653.0 & 13.64 & 13.05 & 0.06 & 0.08 & 0.03 & ... & ...  & Y0YY & CS    & ...     \\
J162802.10-240844.9 & 16.03 & 15.38 & 0.26 & 0.25 & 0.11 & ... & ...  & YYYY & HS    & ...     \\
J162804.54-243448.6 & 15.11 & 12.67 & 0.11 & 0.14 & 0.08 & -54 & II/III& YY0Y & AD    & ...     \\
J162811.08-242919.7 & 11.26 & 10.63 & 0.18 & 0.02 & 0.18 & ... & III  & YYYY & ...   & ...     \\
J162811.46-241458.7 & 15.88 & 15.28 & 0.10 & 0.11 & 0.07 & ... & ...  & Y0Y0 & HS    & ...     \\
J162812.71-241135.7 & 12.09 & 11.07 & 0.14 & 0.21 & 0.14 & -45 & II   & Y0YY & AD    & ...     \\
J162814.72-242846.5 & 15.36 & 14.99 & 0.58 & 0.65 & 0.97 & -31 & ...  & YYYY & AD    & ...     \\
J162814.76-242322.5 & 11.18 & 10.40 & 0.34 & 0.26 & 0.09 & ... & II/III& YYYY & HS    & ...     \\
J162816.41-240738.4 & 14.24 & 13.52 & 0.13 & 0.10 & 0.10 & ... & ...  & YYYY & HS    & ...     \\
J162817.73-241705.1 & 13.23 & 12.30 & 0.10 & 0.15 & 0.06 & -61 & II   & Y0YY & AD    & ...     \\
J162819.82-240906.3 & 12.42 & 11.89 & 0.30 & 0.37 & 0.47 & -35 & III  & YYYY & AD    & ...     \\
J162821.73-241711.3 & 17.20 & 16.63 & 0.28 & 0.34 & 0.47 & -33 & ...  & Y00Y & AD    & ...     \\
J162822.10-241229.6 & 16.69 & 15.96 & 0.32 & 0.33 & 0.41 & ... & ...  & 0YY0 & ...   & ...     \\
J162824.38-242145.3 & 14.93 & 14.34 & 0.11 & 0.11 & 0.03 & ... & III  & Y0YY & CS    & ...     \\
J162838.34-242637.0 & 11.12 & 10.70 & 0.13 & 0.02 & 0.14 & ... & III  & YY0Y & ...   & ...     \\
J162841.53-243323.8 & 11.13 & 10.69 & 0.08 & 0.02 & 0.09 & ... & III  & YY0Y & HS    & ...     \\
J162844.28-244724.7 & 15.92 & 15.42 & 0.17 & 0.16 & 0.06 & ... & ...  & Y0YY & HS    & ...     \\
J162845.88-240835.6 & 13.98 & 13.33 & 0.22 & 0.23 & 0.15 & ... & III  & YYYY & ...   & ...     \\
J162847.02-242813.9 & 12.74 & 12.07 & 0.63 & 0.44 & 0.21 & 65  & II   & YYYY & ...   & ...     \\
J162847.38-242337.2 & 14.17 & 13.91 & 0.07 & 0.11 & 0.05 & -65 & ...  & Y0Y0 & CS,AD & ...     \\
J162847.61-242536.3 & 12.92 & 12.64 & 0.07 & 0.12 & 0.05 & -64 & III  & Y0Y0 & AD    & ...     \\
J162848.69-243150.1 & 14.34 & 14.04 & 0.23 & 0.18 & 0.09 & ... & ...  & YYYY & HS    & ...     \\
J162849.98-243518.1 & 12.44 & 10.85 & 0.06 & 0.05 & 0.03 & ... & III  & Y0YY & CS    & ...     \\
J162850.84-242832.3 & 11.77 & 11.46 & 0.08 & 0.02 & 0.08 & ... & III  & Y0Y0 & HS    & ...     \\
J162852.73-240926.2 & 15.67 & 15.01 & 0.07 & 0.09 & 0.05 & ... & ...  & Y0Y0 & HS    & ...     \\
J162855.25-244515.6 & 17.87 & 17.33 & 0.82 & 0.36 & 0.89 & ... & ...  & YYYY & ...   & ...     \\
J162856.53-244458.3 & 17.75 & 17.30 & 0.53 & 0.55 & 0.72 & -35 & ...  & Y0YY & AD    & ...     \\
J162856.79-243810.1 & 11.16 & 10.53 & 0.13 & 0.01 & 0.14 & -4  & III  & YYYY & AD    & ...     \\
J162856.94-243109.8 & 11.51 & 10.88 & 0.07 & 0.06 & 0.04 & ... & II   & Y0Y0 & CS    & ...     \\
J162857.78-242804.6 & 11.08 & 10.77 & 0.12 & 0.02 & 0.13 & ... & III  & YYYY & ...   & ...     \\
J162904.32-242125.7 & 16.49 & 13.95 & 0.10 & 0.05 & 0.06 & ... & III  & Y00Y & HS    & ...     \\
J162911.85-243955.9 & 16.38 & 15.32 & 0.14 & 0.10 & 0.09 & ... & ...  & Y0YY & HS    & ...     \\
J162915.63-240958.5 & 16.94 & 16.26 & 0.32 & 0.25 & 0.43 & ... & ...  & YYYY & ...   & ...     \\
J162916.32-243844.0 & 14.04 & 13.49 & 0.30 & 0.28 & 0.03 & 82  & ...  & YYYY & CS    & ...     \\
J162919.97-240926.3 & 14.69 & 14.28 & 0.05 & 0.06 & 0.02 & -67 & ...  & Y0Y0 & CS,AD & ...     \\
J162923.28-240914.9 & 13.97 & 13.51 & 0.05 & 0.07 & 0.07 & ... & III  & Y0Y0 & HS    & ...     \\
J162924.58-240332.4 & 14.48 & 14.08 & 0.14 & 0.08 & 0.07 & 42  & ...  & Y00Y & HS    & ...     \\
J162927.18-241020.0 & 17.06 & 16.61 & 0.25 & 0.45 & 0.53 & -35 & ...  & YY0Y & AD    & ...     \\
J162930.31-244139.0 & 11.25 & 10.55 & 0.08 & 0.01 & 0.09 & -7  & III  & Y00Y & AD    & ...     \\
J162931.01-243300.1 & 15.47 & 15.24 & 0.12 & 0.14 & 0.15 & ... & ...  & Y0Y0 & ...   & ...     \\
J162932.80-243305.9 & 11.17 & 10.90 & 0.07 & 0.06 & 0.03 & ... & III  & Y0YY & CS    & ...     \\
J162937.23-243426.8 & 14.19 & 13.90 & 0.23 & 0.25 & 0.21 & ... & III  & YYYY & ...   & ...     \\
J162938.15-243618.4 & 17.21 & 16.87 & 0.41 & 0.32 & 0.49 & ... & ...  & YY0Y & ...   & ...     \\

\footnotetext[1]{J2000.0 IAU designation.}
\footnotetext[2]{Average magnitude over all available epochs.}
\footnotetext[3]{Peak-to-peak magnitude amplitude variation.}
\footnotetext[4]{Slopes in the colour-magnitude diagram \emph{K}~vs.~\emph{H}$-$\emph{K}. See Sect.~3.4 for details.}
\footnotetext[5]{SED class as defined from the combination of IRAC/Spitzer and WFCAM/UKIRT data. See Sect.~3.5 for details.}
\footnotetext[6]{First digit: star identified as variable from the CC indexes; second digit: star identified as variable from the $\chi^{2}_{\nu}$ statistics; third digit: star is variable in year 1; fourth digit: star is variable in year 2 (a star is a long term variable when both these digits are Y). When each of these conditions is verified it is denoted by Y (as in yes), otherwise the digits are set to 0.}
\footnotetext[7]{CP~=~cool spots; HT~=~hot spots; E~=~extinction; AC~=~accretion discs; FO~=~Fu Orionis. See Sect.~4.2 for details.} 
\footnotetext[8]{$\rho$~Ophiuchi members, according to the following studies: 1.~\citet{Comeron1993}; 2.~\citet{Bontemps2001}; 3.~\citet{Gagne2004}; 4.~\citet{Ozawa2005}; 5.~\citet{Wilking2005}.} 

\end{longtable}
%\end{landscape}

}% End \longtab
%______________________________________________   

\Online

\begin{appendix}
\label{appendix:1}
\section{Other variable stars.}

\begin{table}
\caption{List of variables with variability characteristics consistent with active field M dwarfs. The list could still contain young very low mass objects, possibly members of Ophiuchus.}           % title of Table
\centering                          % used for centering table
\begin{tabular}{c c c c c c}        % centered columns 
\hline            % inserts double horizontal lines
\hline
AOC designation & $\textit{$<$H$>$}$    & $\textit{$<$K$>$}$  &  $\Delta\textit{H}$ &  $\Delta\textit{K}$ & $\Delta(\textit{H}$-$\textit{K})$ \\
~ & (mag)          & (mag)        & (mag)          & (mag)          & (mag)        \\
\hline

J162554.64-244112.8 & 18.20 & 17.70 & 0.51 & 0.33 & 0.52  \\
J162556.76-243001.1 & 17.88 & 16.30 & 0.22 & 0.23 & 0.21  \\
J162609.95-240046.7 & 17.22 & 16.84 & 0.10 & 0.27 & 0.32  \\
J162611.05-241208.8 & 17.66 & 17.18 & 0.19 & 0.40 & 0.37  \\
J162612.86-241244.1 & 17.71 & 17.01 & 0.22 & 0.29 & 0.16  \\
J162619.42-243233.8 & 18.38 & 16.98 & 0.35 & 0.31 & 0.32  \\
J162621.54-241242.5 & 18.61 & 17.80 & 0.60 & 0.32 & 0.48  \\
J162623.33-242351.4 & 18.09 & 16.47 & 0.57 & 0.30 & 0.52  \\
J162631.14-241208.5 & 16.91 & 15.65 & 0.12 & 0.11 & 0.12  \\
J162647.88-240946.8 & 16.42 & 15.64 & 0.27 & 0.26 & 0.16  \\
J162700.96-241230.6 & 18.38 & 17.72 & 0.40 & 0.30 & 0.48  \\
J162705.45-240519.9 & 17.02 & 16.43 & 0.20 & 0.15 & 0.13  \\
J162749.40-242919.3 & 17.82 & 16.61 & 0.36 & 0.20 & 0.41  \\
J162749.80-243950.1 & 16.94 & 16.02 & 0.39 & 0.28 & 0.18  \\
J162750.41-244042.7 & 17.36 & 16.43 & 0.27 & 0.07 & 0.26  \\
J162800.15-241222.8 & 18.35 & 17.25 & 0.58 & 0.16 & 0.53  \\
J162801.81-241305.0 & 17.32 & 16.44 & 0.25 & 0.12 & 0.33  \\
J162803.22-240536.7 & 16.18 & 15.63 & 0.32 & 0.20 & 0.15  \\
J162814.41-244546.2 & 16.88 & 16.49 & 0.21 & 0.18 & 0.08  \\
J162818.95-241319.1 & 17.31 & 16.68 & 0.20 & 0.26 & 0.15  \\
J162840.66-243148.3 & 17.72 & 17.22 & 0.42 & 0.32 & 0.35  \\
J162841.89-242705.6 & 16.33 & 16.02 & 0.18 & 0.20 & 0.16  \\
J162844.20-240115.1 & 17.75 & 17.21 & 0.32 & 0.28 & 0.33  \\
J162846.90-240212.9 & 16.82 & 16.46 & 0.08 & 0.32 & 0.28  \\
J162852.34-242849.1 & 17.63 & 17.37 & 0.25 & 0.25 & 0.34  \\
J162858.30-243544.2 & 17.50 & 16.60 & 0.22 & 0.17 & 0.08  \\
J162902.71-242837.7 & 16.12 & 15.78 & 0.21 & 0.13 & 0.21  \\
J162905.65-242314.9 & 17.79 & 17.50 & 0.22 & 0.20 & 0.36  \\
J162911.02-242240.0 & 17.88 & 17.50 & 0.32 & 0.20 & 0.22  \\
J162919.04-241038.8 & 17.72 & 17.15 & 0.27 & 0.24 & 0.22  \\
J162920.93-241936.8 & 16.36 & 15.90 & 0.14 & 0.11 & 0.11  \\
J162929.44-243426.3 & 18.00 & 17.65 & 0.57 & 0.43 & 0.59  \\
J162930.59-241907.8 & 18.21 & 17.68 & 0.42 & 0.43 & 0.28  \\
J162933.26-243359.0 & 17.96 & 17.42 & 0.47 & 0.27 & 0.59  \\
J162937.02-243035.2 & 16.75 & 16.40 & 0.13 & 0.19 & 0.10  \\
J162939.40-242309.4 & 17.19 & 16.82 & 0.30 & 0.19 & 0.21  \\
J162943.28-243056.4 & 17.96 & 17.57 & 0.54 & 0.52 & 0.38  \\
J162943.49-243102.7 & 16.67 & 16.45 & 0.22 & 0.18 & 0.23  \\
J162945.00-243421.7 & 17.71 & 17.39 & 0.46 & 0.33 & 0.28  \\
		   
\end{tabular}
\end{table}

\end{appendix}

%______________________________________________   
\end{document}